\documentclass[twocolumn,aps,prc,floatfix]{revtex4}
\usepackage{graphicx}
\usepackage{dcolumn}
\usepackage{bm}
\usepackage{mhchem}

\begin{document}

\title{Symmetry energy extracted from the S$\pi$RIT pion data in Sn+Sn systems}

\author{Gao-Chan Yong$^{1,2}$}
\affiliation{$^{1}$Institute of Modern Physics, Chinese Academy of Sciences, Lanzhou 730000, China \\
$^{2}$School of Nuclear Science and Technology, University of Chinese Academy of Sciences, Beijing, 100049, China}

\begin{abstract}

With the improved particular Isospin-dependent Boltzmann-Uehling-Uhlenbeck transport model
(impIBUU) including the nucleon-nucleon short-range correlations, the ratios and yields of $\pi^{-}$ and $\pi^{+}$ in Sn+Sn systems with different asymmetries at 270 MeV/nucleon are studied. It is found that the yields of $\pi^{-}$ and $\pi^{+}$ and their ratios in Sn+Sn systems characterized by different
neutron to proton ratios obtained from the very recent S$\pi$RIT pion data, are quite well described by the model, especially when a soft symmetry energy with $L(\rho_{0})$ = 66.75$\pm$24.75 MeV is used. The calculations also clearly demonstrate that the high-momentum tail of the nucleon initialization in momentum space strongly affects the yields and ratios of pion production in Sn+Sn systems with different asymmetries near or below threshold. In addition, given many insoluble theoretical uncertainties in transport models, multi-system experimental measurements with various N/Z asymmetries are proposed to extract the symmetry energy less model-dependently by using heavy-ion collisions.

\end{abstract}


\maketitle

\section{Introduction}
The equation of state (EoS) of dense matter impacts the dynamical evolution and the structure of the emerging compact stars, the conditions for nucleosynthesis and the emerging neutrino
spectra \cite{bau2013}. It also provides a unique chance to learn about the QCD thermodynamics that is not yet accessible to theoretical calculations \cite{eosqcd,eosqcd2}. Thus the EoS of dense matter has attracted a lot of attention over the last decades \cite{eosrmp17}. The equation of state (EoS) of
nuclear matter at density $\rho$ and isospin asymmetry
$\delta$ ($\delta =(\rho_n-\rho_p)/(\rho_n+\rho_p)$) is usually
expressed as \cite{esym1991,li08,bar05}
\begin{equation}
E(\rho ,\delta )=E(\rho ,0)+E_{\text{sym}}(\rho )\delta ^{2}+\mathcal{O}%
(\delta ^{4}),
\end{equation}%
where $E_{\text{sym}}(\rho)$ is nuclear symmetry energy.
Currently the EoS of symmetric nuclear matter $E(\rho, 0)$ is relatively well
constrained \cite{pawl2002,eosrmp17} while the EoS of asymmetric nuclear matter, especially the nuclear symmetry energy, is still controversial \cite{Guo14,epja19}. A lot of studies on probing the nuclear symmetry energy have been carried out for many years \cite{npryo2020}.
To constrain the symmetry energy, many terrestrial experiments are being carried out (or planned) using a wide variety of advanced facilities, such as the GSI Facility for Antiproton and Ion Research (FAIR) in Germany \cite{fopi16}, the Cooling Storage Ring on the Heavy Ion Research Facility HIRFL-CSR at IMP in China \cite{csr} and the Facility for Rare Isotope Beams (FRIB) in the Untied States \cite{frib}, the Radioactive Isotope Beam Factory (RIBF) at RIKEN in Japan \cite{shan15} as well as the Rare Isotope Science Project (RISP) in Korea \cite{korea}.
Constraints on the high-density behavior of the symmetry energy can be highly relevant to a series of properties of neutron stars \cite{Lat01,Lat04,Vil04,Ste05}.
Experimentally, related measurements of pion data in the reaction systems $^{132}\rm {Sn}+^{124}\rm {Sn}$, $^{112}\rm {Sn}+^{124}\rm {Sn}$ and $^{108}\rm {Sn}+^{112}\rm {Sn}$ at 270 MeV/nucleon have been done by the S$\pi$RIT collaboration at the Radioactive Isotope Beam Factory operated by the RIKEN Nishina center and CNS, University of Tokyo \cite{spidata2021}.

Pion production in heavy-ion collisions at intermediate energies
has attracted more and more theoretical attention in recent years \cite{ko1,ko2,ko3,pi3,cozma17, un1,un2,un3,un4}, simply because the $\pi^{-}/\pi^{+}$ ratio is a potentially sensitive observable of the symmetry energy \cite{liprl2002} and can be easily measured compared to uncharged neutrons. In the present studies on pion production, we use the improved particular Isospin-dependent Boltzmann-Uehling-Uhlenbeck (impIBUU) transport model \cite{yong20171,yong20151,yongliphoto}. This model mainly includes the nucleon-nucleon short-range correlations in initialization and mean-field potential, the isospin-dependent in-medium elastic and inelastic baryon-baryon cross sections as well as the momentum-dependent isoscalar and isovector nucleon and pion potentials \cite{yong20151,yong20152,yong20153}. With a soft symmetry energy in the practical impIBUU calculations in Sn+Sn systems with different asymmetries, it is found that the output yields and ratios of $\pi^{-}$ and $\pi^{+}$ in these systems fit the recently released S$\pi$RIT pion data very well. The simultaneous reproductions of tri-system Sn+Sn pion experimental measurements by the impIBUU model permit the extraction of a relatively reliable symmetry energy.

\section{Descriptions of the impIBUU transport model}

The impIBUU model originates from the IBUU04 model \cite{lyz05}.
The model describes the time evolution of the single particle phase space distribution function.
In coordinate space, the initial density distributions of neutron and proton in projectile and target nuclei are given by the Skyrme-Hartree-Fock calculations using the Skyrme
M$^{\ast}$ force parameters \cite{skyrme86}. In momentum space, proton and neutron
momentum distributions with a high-momentum tail (HMT) reaching 1.75 times local Fermi momentum are used \cite{yong20151}. The upper limit of 1.75 times fermi momentum of nucleon movement is obtained from the constraints of combining pion data with photon data and transport model calculations \cite{yong20151}.
In practice, a nucleus is divided into many spherical shells centered around its center of mass. The local Fermi momenta of nucleons in each shell of radius $r$ are calculated according to the local Thomas-Fermi approximation
$
 k_{F_{n,p}}(r)= [3\pi^{2}\rho(r)_{n,p}]^{\frac{1}{3}}.
$
In each shell, the nucleon momenta are generated according to the following distributions with HMTs reaching $\lambda k_{F_{n,p}}(r) = 1.75\times k_{F_{n,p}}(r)$ \cite{yong20151,henprc15}
\begin{eqnarray}
n(k)=\left\{%
  \begin{array}{ll}
    C_{1}, & \hbox{$k \leq k_{F}$;} \\
    C_{2}/k^{4}, & \hbox{$k_{F} < k < \lambda k_{F}$}.\\
\end{array}%
\right.
\label{nk}
\end{eqnarray}
Where $C_1$ and $C_2$ are determined by the specified fractions of neutrons and protons in their respective HMTs. They are normalized as
$
\int_{0}^{\lambda k_{F}}n(k)k^{2}dk = 1.
\label{nk1}
$
It is known that for medium and heavy nuclei about 20\% nucleons are in the HMT \cite{e93,e96,sci08}. When adopting the n-p dominance model requiring equal numbers of neutrons and protons in the HMT \cite{sci14}, the fraction of nucleons in the HMT should decrease as asymmetry $\delta$ increases. The fraction of total nucleons in the HMT is assumed to decrease as $20\%(1-\delta^{2})$, thus $10\%(1\mp\delta)$ of neutrons or protons are distributed in their respective HMTs \cite{nature2018}. The rest of them are then distributed in their respective Fermi seas. The nucleon momentum distribution in the nucleus is then formally written as \cite{yongliphoto}
\begin{equation}
 n_{n,p}(k)= \frac{1}{N,Z}\int _{0}^{r_{max}}d^{3}r\rho_{n,p}(r)\cdot n[k,k_{F_{n,p}}(r)],
\end{equation}
with $N$ and $Z$ being the total numbers of neutrons and protons in a nucleus.

In the impIBUU model, the following isospin- and momentum-dependent single-nucleon
potential (MDI) is used \cite{spp1,yong20152}
\begin{eqnarray}
U(\rho,\delta,\vec{p},\tau)&=&A_u(x)\frac{\rho_{\tau'}}{\rho_0}+A_l(x)\frac{\rho_{\tau}}{\rho_0}\nonumber\\
& &+B(\frac{\rho}{\rho_0})^{\sigma}(1-x\delta^2)-8x\tau\frac{B}{\sigma+1}\frac{\rho^{\sigma-1}}{\rho_0^\sigma}\delta\rho_{\tau'}\nonumber\\
& &+\frac{2C_{\tau,\tau}}{\rho_0}\int
d^3\,\vec{p^{'}}\frac{f_\tau(\vec{r},\vec{p^{'}})}{1+(\vec{p}-\vec{p^{'}})^2/\Lambda^2}\nonumber\\
& &+\frac{2C_{\tau,\tau'}}{\rho_0}\int
d^3\,\vec{p^{'}}\frac{f_{\tau'}(\vec{r},\vec{p^{'}})}{1+(\vec{p}-\vec{p^{'}})^2/\Lambda^2},
\label{buupotential}
\end{eqnarray}
where $\rho_0$ denotes saturation density, $\tau, \tau' = 1/2(-1/2)$ for neutron (proton).
The parameter $x$ is introduced to mimic different forms of the symmetry energy predicted by various
many-body theories without changing any
property of the symmetric nuclear matter and the symmetry energy
at normal density. The effects of short-range correlations on the mean field are reflected by the parameters in the single nucleon potential Eq.~(\ref{buupotential}), which can be found in Refs.~\cite{yong20152,yong20171}, i.e., $A_u(x)$ = 33.037 - 125.34$x$
MeV, $A_l(x)$ = -166.963 + 125.34$x$ MeV, B = 141.96 MeV,
$C_{\tau,\tau}$ = 18.177 MeV, $C_{\tau,\tau'}$ = -178.365 MeV, $\sigma =
1.265$, and $\Lambda = 630.24$ MeV/c.
With these settings, the empirical values of nuclear matter at normal density are reproduced, i.e., the saturation density $\rho_{0}$ = 0.16 fm$^{-3}$, the binding energy $E_{0}$ = -16 MeV, the incompressibility $K_{0}$ = 230 MeV \cite{Oertel17, k230}, the isoscalar effective mass
$m_{s}^{*} = 0.7 m$ \cite{pawel00}, the single-particle potential
$U^{0}_{\infty}$ = 75 MeV at infinitely large nucleon momentum at
saturation density in symmetric nuclear matter, the symmetry
energy $E_{\rm sym}(\rho_0) = 34.57$ MeV \cite{chenlw2021,reed2021}.
With these settings, the experimental Hama potential at saturation density \cite{hama90} is well reproduced. For most symmetry-energy sensitive observables, it is the symmetry potential that plays a role. Therefore, the symmetry potential is the most crucial ingredient of a transport model used to study the symmetry energy. It is worth noting here that the momentum dependence of the symmetry potential given by Eq.~(\ref{buupotential}) fits the experimental data very well \cite{yong20152,smp04,changxu10}.
\begin{figure}[th]
\centering
\includegraphics[width=0.5\textwidth]{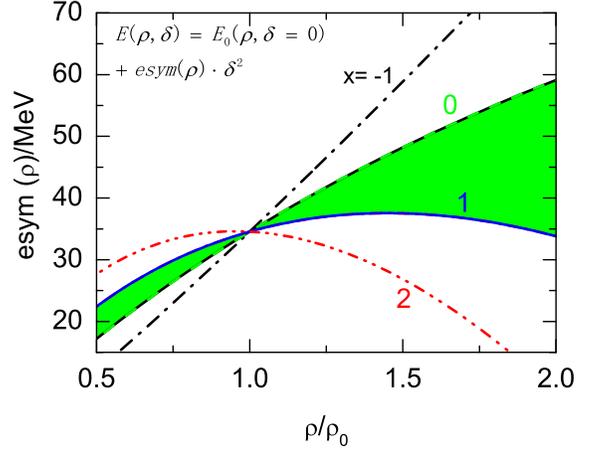}
\caption{The density-dependent symmetry energy of the single particle potential Eq.~(\ref{buupotential}) with different $x$ parameters. The green shade denotes the possible symmetry energy discussed in the text.} \label{esym}
\end{figure}
In order to obtain the density-dependent symmetry energy from Eq.~(\ref{buupotential}), one needs the density-dependent kinetic symmetry energy, which can be obtained from the correlated Fermi gas (CFG) model \cite{henprc15}. Fig.~\ref{esym} shows the corresponding symmetry energy of the single particle potential  Eq.~(\ref{buupotential}) with different $x$ parameters. $x$ = 1, 0, -1 cases respectively correspond to the slopes ($L (\rho_{0}) \equiv 3\rho_{0}dEsym(\rho)/d\rho$) of 42, 91.5, 141.5 MeV while $x$ = 2 corresponds to a negative slope, which seems impossible.

The isospin-dependent baryon-baryon ($BB$) elastic and inelastic scattering cross sections in medium $\sigma_{BB}^{medium}$ are both reduced compared with their free-space value
$\sigma _{BB}^{free}$ by a factor of \cite{yong20152,Lichen05}
\begin{eqnarray}
R^{medium}_{BB}(\rho,\delta,\vec{p})&\equiv& \sigma
_{BB}^{medium}/\sigma
_{BB}^{free}\nonumber\\
&=&(\mu _{BB}^{\ast }/\mu _{BB})^{2},
\end{eqnarray}
where $\mu _{BB}$ and $\mu _{BB}^{\ast }$ are the reduced masses
of the colliding baryon pairs in free space and medium,
respectively. The invariant mass of the two-particle system is presently not modified
since at low beam energies pion production generally comes from multiple scatterings of among particles
and the ingoing invariant mass in medium is equal to that of outgoing state in medium is assumed \cite{mosel03}.
The effective mass of baryon in isospin asymmetric nuclear matter
is expressed as
$
\frac{m_{B}^{\ast }}{m_{B}} = 1/(1+\frac{m_{B}}{p}\frac{%
dU}{dp}).
$
For the resonance, its free mass is determined according to a modified Breit-Wigner
function \cite{dmass91,yong20171}, its single particle potential $U$ is divided into the single particle potentials of neutron and proton according to Clebsch-Gordon coefficients for isospin coupling \cite{linpa2002}. As argued in Ref.~\cite{delta15}, the resonance potential in fact has negligible effects on the charged pion ratio especially when the symmetry energy is mildly soft.

In the impIBUU model, the specific pion production mechanism is via $\Delta$ resonance model Ref.~\cite{yong20171,ono19}. We actually do not propagate the full spectral function of pions \cite{pionp1,pionp2,pionp3,pionp4,pionp5}. We describe the particles as classical quasi-particles by adding an effective optical potential for the pions in nuclear medium. A density- and momentum-dependent pion potential including isoscalar and isovector contributions is used \cite{pionp6,yong20153}. It is repulsive at low pionic momenta but attractive at high pionic momenta. The isoscalar potential is overall positive but the isovector potential is positive for $\pi^{-}$ while negative for $\pi^{+}$. In the processes relevant to resonance production and absorption, since the total gain and loss of the potential energies caused by different mean-field potentials of nucleons and $\Delta$ resonance is cancelled out,
the global energy conservation is kept \cite{cozma17,ko2}. Since a single reaction channel always has energy exchange with surrounding particles, the single energy conservation is abandoned thus not modified.

Overall, the distinguishing feature of the present impIBUU model is its consideration of the effects of the nucleon-nucleon short-range correlations. With such consideration, the nucleon momentum initialization in colliding nuclei, the kinetic symmetry energy as well as its mean-field potential are all very different from those of the transport models used to decode the density-dependent symmetry energy from the S$\pi$RIT pion data in Sn+Sn systems as shown in Ref.~\cite{spidata2021}.

\section{Results and Discussions}

By using the above impIBUU model, the yields and ratios of $\pi^{-}$ and $\pi^{+}$ in $^{132}\rm {Sn}+^{124}\rm {Sn}$, $^{112}\rm {Sn}+^{124}\rm {Sn}$ and $^{108}\rm {Sn}+^{112}\rm {Sn}$ reactions at 270 MeV/nucleon with an impact parameter b = 3 fm are calculated \cite{spidata2021}. In the calculations, only the symmetry energy stiffness parameter $x$ is varied. In order to see the effects of the HMT on pion production, we also turn off the HMT in the calculations.
\begin{figure}[th]
\centering
\includegraphics[width=0.5\textwidth]{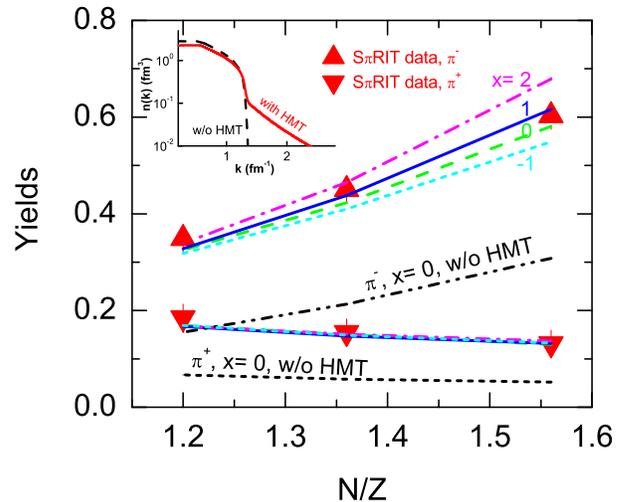}
\caption{Charged pion yields as a function of N/Z for $^{132}\rm {Sn}+^{124}\rm {Sn}$ (N/Z = 1.56), $^{112}\rm {Sn}+^{124}\rm {Sn}$ (N/Z = 1.36) and $^{108}\rm {Sn}+^{112}\rm {Sn}$ (N/Z = 1.2) reactions at 270 MeV/nucleon with different symmetry energies. The effects of the HMT on the charged pion yields are shown for $x$ = 0 case. The inserted plots show nucleon momentum distributions in nucleus with or without the HMT. Data are taken from Ref.~\cite{spidata2021}.} \label{yd}
\end{figure}
Fig.~\ref{yd} shows the numbers of charged pions produced in Sn+Sn systems with different N/Z asymmetries. By comparison, it is seen that our results on pion yields fit the experimental S$\pi$RIT pion data \cite{spidata2021} quite well, especially with the symmetry energy stiffness parameters $x$ = 1, 0. The yields of $\pi^{+}$ are not sensitive to the symmetry energy while the yields of $\pi^{-}$ are very sensitive to the symmetry energy especially for larger N/Z system. This is consistent with that shown in Ref.~\cite{lyz05}, simply because positively charged particles also suffer from the Coulomb potential. Because $\pi^{-}$'s ($\pi^{+}$'s) are mainly from n-n (p-p) collisions, one thus sees more $\pi^{-}$'s being produced in neutron-rich systems. As asymmetry N/Z increases, more neutrons are involved into scatterings with neutrons or protons. Thus more n-n collisions produce more $\pi^{-}$'s, more p-n collisions produce more $\pi^{0}$'s, and less p-p collisions produce less $\pi^{+}$'s (n-p colliding cross section is generally larger than that of p-p, thus proton trends to collide with neutron rather than proton). One thus sees $\pi^{-}$ ($\pi^{+}$) yield increases (decreases) monotonously as N/Z increases. 
Nucleon's high-momentum tail trends to cause a certain proportion of nucleon-nucleon collisions with center of mass energy reaching or exceeding the pion production threshold, therefore the yields of pion with the HMT are evidently higher than those without the HMT. From Fig.~\ref{yd}, it is also seen that the effects of the HMT on charged pion yields can be as high as 50\%. Since the short-range correlations and the HMT of the nucleon momentum distribution in the nucleus have been experimentally confirmed \cite{sci14}, one should involve such physics into a transport model. From the $\pi^{-}$ production in Fig.~\ref{yd}, it is seen that the symmetry energy with $x$ = 1, or $x$ = 0 is supported by the S$\pi$RIT pion data.

\begin{figure}[th]
\centering
\includegraphics[width=0.53\textwidth]{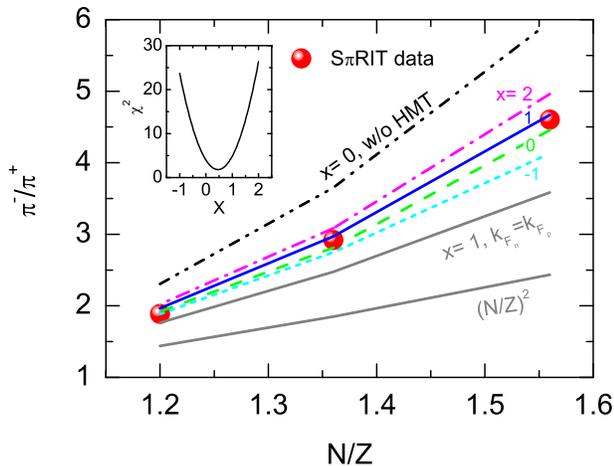}
\caption{Charged pion yield ratios as a function of N/Z for Sn+Sn systems with different symmetry energies. The effects of the HMT on the charged pion yield ratios are shown for $x$ = 0 case. The gray line (N/Z)$^{2}$ denotes the prediction of $\Delta$ resonance model while $k_{F_{n}} = k_{F_{p}}$ stands for equality of fermi momenta of neutrons and protons. Data are taken from Ref.~\cite{spidata2021}.} \label{sg}
\end{figure}
To reduce systematic errors, instead of using charged pion yields, one always analyzes pion yield ratios.
Fig.~\ref{sg} shows the charged pion yield ratios as a function of N/Z. Because the yields of $\pi^{+}$ ($\pi^{-}$) monotonously decrease (increase) with N/Z as shown in Fig.~\ref{yd}, the ratio of yields of $\pi^{-}$ and $\pi^{+}$ increases with N/Z. It is seen that the calculated ratios of $\pi^{-}/\pi^{+}$ with $x$ = 1 and 0 overall fit experimental data quite well whereas the results with $x$ = 2, -1 deviate from the data. The results without the HMT deviate the data evidently. Because nucleon-nucleon short-range correlations speed up protons more evidently than neutrons in neutron-rich matter \cite{nature2018}, energies of proton-proton collisions become larger. Therefore more $\pi^{+}$'s are produced. Without the HMT, one would see an opposite behavior. Thus without the HMT, the ratios of yields of $\pi^{-}$ and $\pi^{+}$ are evidently higher than those with the HMT.

In Fig.~\ref{sg}, the gray line (N/Z)$^{2}$ denotes the prediction of $\Delta$ resonance model for pion production \cite{spidata2021,fopi2007,liprl2002,stock86}. It is seen that this prediction is far below the results given by the impIBUU model as well as the experimental data. Since for sub-threshold pion production, nucleon initial fermi momentum plays a more important role than at high energies, the values of $\pi^{-}/\pi^{+}$ ratio are clearly reduced when setting $k_{F_{n}} = k_{F_{p}}$ in the momentum initialization. The larger neutron fermi momentum $k_{F_{n}} > k_{F_{p}}$ in neutron-rich systems causes more $\pi^{-}$ than $\pi^{+}$ to be produced. The ratio of $\pi^{-}/\pi^{+}$ is thus higher than that from the (N/Z)$^{2}$ prediction of $\Delta$ resonance model, especially for larger N/Z system. From the above demonstrations of the effects of the HMT and the neutron and proton fermi momentum settings on the $\pi^{-}/\pi^{+}$ ratio, one can deduce that the initial nucleon momentum distribution really plays an important role on pion production at the incident beam energy of 270 MeV/nucleon. Note here that the HMT is different from nucleon momentum distribution representing high temperatures that quickly develop during the collision, at least the n/p ratios from their respective high-momentum distributions are different.

\begin{figure}[th]
\centering
\includegraphics[width=0.5\textwidth]{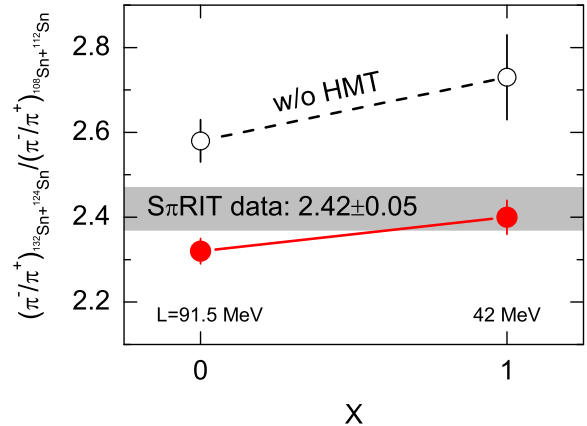}
\caption{Double pion yield ratios for $^{132}\rm {Sn}+^{124}\rm {Sn}$ and $^{108}\rm {Sn}+^{112}\rm {Sn}$ reactions at 270 MeV/nucleon with different symmetry energies. The effects of the HMT on the double pion yield ratios are also shown. Data are taken from Ref.~\cite{spidata2021}.} \label{dr}
\end{figure}
To further reduce systematic errors, it is more attractive to study the double $\pi^{-}/\pi^{+}$ ratio from neutron-rich and neutron-deficient systems \cite{yongdb06,guodb14,spidata2021}, because one generally considers that such double ratio from neutron-rich and neutron-deficient systems can largely reduce uncertainties such as the Coulomb interactions, unknown in-medium nucleon-nucleon scattering cross sections and some isospin-independent uncertainties while keeping the effects of the symmetry energy. Fig.~\ref{dr} shows the double $\pi^{-}/\pi^{+}$ ratio for $^{132}\rm {Sn}+^{124}\rm {Sn}$ and $^{108}\rm {Sn}+^{112}\rm {Sn}$ systems. Since from Fig.~\ref{yd} and Fig.~\ref{sg}, the S$\pi$RIT pion data favor the symmetry energies with $x$ = 1 and 0, in Fig.~\ref{dr}, the double $\pi^{-}/\pi^{+}$ ratios are shown only for the symmetry energy parameters $x$ = 1 and 0. By comparison, it is seen that the result with the symmetry energy parameter $x$ = 1 fits the double $\pi^{-}/\pi^{+}$ ratio data quite well. The result with $x$ = 0 is somewhat lower than the data. As expected, without the HMT, the double $\pi^{-}/\pi^{+}$ ratios given by the transport model are evidently higher than the data.

Because the two systems are in fact both neutron-rich, sensitivity of the double $\pi^{-}/\pi^{+}$ ratio to the symmetry energy is reduced to some extent. Systematic errors of the double $\pi^{-}/\pi^{+}$ ratio for the two neutron-rich systems from uncertainties of the undetermined momentum-, density- and asymmetry-dependent in-medium nucleon-nucleon elastic/inelastic scattering cross sections \cite{inmed2021} and the momentum-, density-dependent symmetry potential in fact cannot be cancelled out by such operation. For these reasons, to get more reliable constraints on the symmetry energy, it is more powerful to carry out a multi-system comparison of experimental measurements and theoretical simulations, such as, a series of systems with N/Z = 1, 1.2, 1.4, 1.6, etc. One uses the system with N/Z = 1 as benchmark or starting point to study any symmetry-energy related observable, to check/correct the symmetry-energy-independent parts of the model; The system with the largest asymmetry N/Z = 1.6 can be used to ``search'' the form of the symmetry energy; Systems with N/Z = 1.2, 1.4 are used to check/ensure the ``correctness'' of the form of the symmetry energy. If the model cannot simultaneously fit the data from two or more systems (N/Z = 1.2, 1.4,...) by using the same symmetry energy, this means the effects of the \emph{deviated} isospin-dependent (but non-symmetry-energy dependent) parts of the transport model (such as the isospin-dependent HMT and the isospin-dependent in-medium inelastic cross section) are not properly cancelled out mutually, one has to adjust the model to fit three or more system measurements. For a single neutron-rich system, when the HMT reduces the $\pi^{-}/\pi^{+}$ ratio, one can adjust the stiffness of the symmetry energy to enhance the ratio of the $\pi^{-}/\pi^{+}$ again. But for two or more systems with different N/Z asymmetries, the same stiffness of the symmetry energy cannot cancel out the HMT since the HMT and the symmetry energy have different behavior as a function of the N/Z asymmetry. The basic philosophy of multi-system measurements with different N/Z asymmetries is that nothing has the same behavior as the symmetry energy with increase of N/Z asymmetry, and thus cannot be substituted by some kind of factor. As the number of systems with different N/Z increase, the symmetry energy fitting all the data would be the right form. Given many unresolved uncertainties especially some crucial inputs of transport model, multi-system measurement may be an alternative/practical way to unveil the veil of the symmetry energy for the foreseeable future.

The Sn+Sn tri-system measurements of the S$\pi$RIT pion experiments with different asymmetries in fact basically meet the above criteria. System $^{108}\rm {Sn}+^{112}\rm {Sn}$ roughly acts as the symmetric system, can be used to check/correct the symmetry-energy-independent parts of the model. System $^{132}\rm {Sn}+^{124}\rm {Sn}$ is used to ``search'' the form of the symmetry energy and system $^{112}\rm {Sn}+^{124}\rm {Sn}$, ensures the ``correctness'' of the extracted symmetry energy. For this reason, although the double ratio, as shown in Fig.~\ref{dr}, can reduce some systematic errors, it is more powerful to fit tri-system Sn+Sn measurements to extract the symmetry energy. As shown in Fig.~\ref{sg}, the agreements of the simulated $\pi^{-}/\pi^{+}$ ratios with the tri-system Sn+Sn measurements indicate our extracted symmetry energy ($0 < x < 1$ or $L(\rho_{0})$ = 66.75$\pm$24.75 MeV) is reliable. This result is also consistent with the recent constraints of combining astrophysical data with PREX-II and chiral effective field theory \cite{reed2021}.

The inconsistent conclusions while decoding previous FOPI Au+Au data \cite{fopi2007} by different transport models \cite{un1,un2,un3,un4} originate from different physical inputs and computational methods used in different models. Such conflicting conclusions could be cured through multi-system comparisons as discussed above, since a single-system experimental measurement is always easily reproduced by any transport model. Putting some uncertainties as frequently mentioned in the literature aside, in case all the yields and ratios of $\pi^{-}$ and $\pi^{+}$ given by the seven transport models can reproduce all the experimental data in Sn+Sn systems with \emph{three} asymmetries \cite{spidata2021}, a roughly consistent form of the symmetry energy is expected to be achieved.

\section{Conclusions}

In summary, within the framework of the impIBUU model, we decode the recently released S$\pi$RIT pion data in Sn+Sn systems. It is found that a soft symmetry energy with the slope of $L(\rho_{0})$ = 66.75$\pm$24.75 MeV is favored. Nucleon momentum initial distribution such as the HMT in projectile and target nuclei significantly affects the single and double $\pi^{-}/\pi^{+}$ ratios, thus evidently affects the extraction of the symmetry energy from experimental data. Given many insoluble theoretical uncertainties in transport model, multi-system (with N/Z = 1, 1.2, 1.4, 1.6,...) measurement and comparison may be an alternative/practical way to reliably extract the symmetry energy from heavy-ion collisions.

The author would like to express his gratitude to Prof. Dr. Wolfgang Trautmann for carefully reviewing the manuscript. This work is supported in part by the National Natural Science Foundation of China under Grant No. 11775275.

\end{document}